\documentclass[nojss]{jss}
\usepackage[T1]{fontenc}
\usepackage[latin9]{inputenc}
\usepackage{amsthm}
\usepackage{amsmath}
\usepackage{verbatim}

\makeatletter
\numberwithin{equation}{section}
\theoremstyle{plain}
\newtheorem{thm}{\protect\theoremname}[section]
\theoremstyle{plain}
\newtheorem{algorithm}[thm]{\protect\algorithmname}

\makeatother

\usepackage[USenglish]{babel}
\providecommand{\algorithmname}{Algorithm}
\providecommand{\theoremname}{Theorem}

\DeclareMathOperator*{\argmax}{arg\,max}

\author{Matthew Friedlander \\ Tanenbaum-Lunenfeld Research Institute \And
              Adrian Dobra \\ University of Washington \AND
              H\'el\`ene Massam \\ York University \And
              Laurent Briollais \\ Tanenbaum-Lunenfeld Research Institute}
\title{Analyzing Genome-wide Association Study Data with the \proglang{R} Package  \pkg{genMOSS} }

\Plainauthor{M. Friedlander, A. Dobra, H. Massam, L. Briollais} 
\Plaintitle{} 
\Shorttitle{The \pkg{genMOSS} \proglang{R} Package}

\Abstract{
The \proglang{R} package \citep{R} \pkg{genMOSS} is specifically designed for the Bayesian analysis of genome-wide association study data. The package implements the mode oriented stochastic search procedure of \citep{Dobra2} as well as a simple moving window approach to identify combinations of single
nucleotide polymorphisms associated with a response. The prior used in Bayesian computations is the generalized hyper Dirichlet of \citep{Massam}.
}

\Plainkeywords{keywords, comma-separated, not capitalized} 

\Address{

  Matthew Friedlander\\
  Tanenbaum-Lunenfeld Research Institute\\
  Mount Sinai Hospital\\
  Toronto, ON\\
  Canada\\
  E-mail: \email{friedlander@lunenfeld.ca}\\

    Adrian Dobra\\
  Department of Statistics\\
  University of Washington\\
  Seattle, WA\\
  USA\\
  E-mail: \email{adobra@u.washington.edu}\\

  H\'el\`ene Massam\\
  Department of Mathematics and Statistics\\
  York University\\
  Toronto, ON\\
  Canada\\
  E-mail: \email{massamh@yorku.ca}\\
  
  Laurent Briollais\\
  Tanenbaum-Lunenfeld Research Institute\\
  Mount Sinai Hospital\\
  Toronto, ON\\
  Canada\\
  E-mail: \email{laurent@lunenfeld.ca}\\
  URL: \url{http://www.lunenfeld.ca/researchers/briollais}\\
}

\begin{document}
\section{Introduction}

Genome-wide association studies (GWAS) produce large amounts of biological information that is used for the phenotyping of many diseases. A typical GWAS
dataset can have thousands of potential predictive single nucleotide-polymorphisms (SNPs) and the aim is to find a small subset of these predictors that are
related to a particular disease. Many variable selection techniques use univariate tests that individually measure the dependency between each candidate
predictor and the response - see, for example, \citep{Golub, Nguyen, Dudoit, Tusher}. With univariate testing there are complex issues related to assessing the
statistical significance of a large number of null hypothesis tests \citep{Benjamini, Efron, Storey}. Moreover, as \citep{Schaid} points out, single marker
analysis exploits only a fraction of the information available. The alternative is to take into consideration combinations of predictors, which leads to
an exponential increase in the number of candidate models. One simple approach is to group SNPs together in sequence over a moving window and examine their
association with the response. This can help identifying particular genetic regions of interest. A weakness of this approach is the subjective selection of the
window size and the inability to capture multi-SNP effects among SNPs in different regions of the genome \citep{Sun, Wu}. To allow any potential combinations of SNPs to be selected, stochastic search algorithms are necessary that are capable of quickly finding the SNPs most associated with the disease of interest.

The mode oriented stochastic search (MOSS) algorithm is a two-stage Bayesian variable selection procedure that aims to identify combinations of SNPs that are
associated with a response. If we let $Y$ be a response (e.g. disease status) and $X$ be a set of predictors, the first stage is to maximize $P(Y|X)$. The rationale is that if there
is a strong relationship between $Y$ and $X$ then the probability of the regression $Y|X$ should be relatively high. The second stage is to search the space of hierarchical log-linear models to identify 
the most relevant interactions among the variables in each of the top regressions. By using the generalized hyper Dirichlet prior of \citep{Massam}, the computations in both steps can be done efficiently. 
Once a set of promising log-linear models has been found (at the end of stage two), model averaging can be used to build a classifier for predicting the response. 
The efficacy of the classifier can be assessed using k-fold cross validation. 

The main objective of this paper is to describe the \pkg{genMOSS} \proglang{R} package which is a free implementation of the MOSS and moving window approaches for the Bayesian
analysis of GWAS data. In section 2, we review the MOSS procedure which is described in detail in \citep{Dobra1}. In section 3, we briefly discuss the moving
window approach. In section 4, we illustrate the use of the package with some examples.

\section{MOSS}

Suppose that $V$ is a set of classification criteria. Let $Y=X_{\gamma},\gamma\in V$ be a response variable and $X_{A},A\subset V\backslash\gamma$ be a set of explanatory variables. In the first stage of MOSS we assume that the saturated log-linear model holds for all variables in $V$. After collapsing over
any subset of variables in $V$ we still retain a saturated model. To see this, consider a simple example with the saturated log-linear model for three binary variables $X,Y,$ and $Z$:
\[
\log p_{ijk}=\theta+\theta_{i}^{X}+\theta_{j}^{Y}+\theta_{k}^{Z}+\theta_{ij}^{XY}+\theta_{ik}^{XZ}+\theta_{jk}^{YZ}+\theta_{ijk}^{XYZ}.
\]
The marginal probability $p_{ij+}=p_{ij1}+p_{ij2}$ satisfies:
\[
p_{ij+}=\exp\left(\theta+\theta_{i}^{X}+\theta_{j}^{Y}+\theta_{ij}^{XY}\right)\sum_{k=1}^{2}\exp\left(\theta_{k}^{Z}+\theta_{ik}^{XZ}+\theta_{jk}^{YZ}+\theta_{ijk}^{XYZ}\right).
\]
By taking logarithms we have:
\[
\log p_{ij+}=\theta+\theta_{i}^{X}+\theta_{j}^{Y}+\left\{
\theta_{ij}^{XY}+\log\sum_{k=1}^{2}\exp\left(\theta_{k}^{Z}+\theta_{ik}^{XZ}+\theta_{jk}^{YZ}+\theta_{ijk}^{XYZ}\right)\right\} .
\]
The term in the braces is a function only of $i$ and $j$ so that replacing it with $\lambda_{ij}$ we have the saturated model:
\[
\log p_{ijk}=\theta+\theta_{i}^{X}+\theta_{j}^{Y}+\lambda_{ij}^{XY}.
\]
Our aim in the first step of the MOSS procedure is to search for sets $A$ such that the probability, or marginal likelihood, of the regression $r=Y|X_{A}$ 
\begin{equation}
P(r)=P(Y|X_{A})=\frac{P(Y,X_{A})}{P(X_{A})}
\label{2.1}
\end{equation}
is highest. Sample sizes in GWAS data are typically small compared to the number of variables, so $A$ should contain a small number of SNPs, say 2-5. We note that \ref{2.1} is the ratio of the
marginal likelihoods of two saturated log-linear models. By putting the generalized hyper Dirichlet conjugate prior of \citep{Massam}
on the log-linear parameters, an explicit formula exists for \ref{2.1} making its computation particularly easy. We now give the general algorithm to search for
the top regressions in terms of \ref{2.1}.

Let $R$ denote a set of possible regression models. We associate with each candidate model $r\in R$ a neighbourhood $\textrm{nbd}(r)\subset R$. Any two models
$r,r'\in R$ are connected through a path $r=r_{1},r_{2},...,r_{l}=r'$ such that $r_{j}\in\textrm{nbd}(r_{j-1})$ for $j=2,...,l$. The neighbourhood of
$r=Y|X_{A}$ is obtained by addition moves, deletion moves, and replacement moves. In an addition move, we individually include in $A$ any variable in
$V\backslash A$. In a deletion move, we individually delete any variable that belongs to $A$. For a replacement move, we individually replace any one variable
in $A$ with any one variable in $V\backslash A$. The first stage of the MOSS procedure is as follows. 

\begin{algorithm}
\normalfont
We make use of a current list of regressions $S$ that is updated during the search. Define
\[
S(c)=\left\{ r\in S:P(r)\ge c\max_{r'\in R}P(r')\right\}
\]
where $c\in(0,1)$. A regression $r\in S$ is called explored if all of its neighbours $r'\in\textrm{nbd}(r)$ have been visited.
\end{algorithm}

\begin{enumerate}
\item Initialize a starting list of regressions $S$. For each $r\in S$, calculate and record its marginal likelihood $P(r)$. Mark $r$ as unexplored.
\item Let $L$ be the set of unexplored regressions in $S$. Sample an $r\in L$ according to probabilities proportional with $P(r)$ normalized within $L$. Mark
$r$ as unexplored.
\item For each $r'\in\textrm{nbd}(r)$, check if $r'$ is currently in $S$. If it is not, evaluate and record its marginal likelihood $P(r').$ Eliminate the
regressions $S\backslash S(c')$ for some pre-chosen value $0<c'<c.$
\item With probability $q$ eliminate from $S$ the regressions in $S\backslash S(c).$
\item If all the regressions in $S$ are explored STOP. Otherwise return to step 2.
\end{enumerate}

The role of the parameters $c,c',$ and $q$ is to limit the number of regressions that need to be visited to a manageable number. It is recommended to run the
algorithm with different choices of these quantities to determine the sensitivity of the models selected. However, the default values supplied with the package
have worked well for many datasets.

At the end of the first stage we will have a set of top regressions each involving a small number of variables. At this point, we relax the assumption that the
saturated model holds for all the variables $V$. In the second stage, we search the space of hierarchical log-linear models to identify the most relevant
interactions among the variables in each regression. We do a separate search for each regression looking for the hierarchical log-linear model $m$ with the
highest marginal likelihood. If we let $t_{m}$ denote the sufficient statistic for the log-linear parameters in $m$ and $M$ denote the space of models, then we
seek to find

\begin{equation}
\argmax_{m\in M}P\left(t_{m}|m\right).
\label{2.2}
\end{equation}

To do this, we once again begin by defining the concept of a neighbourhood. The neighbourhood of a hierarchical model $m$ consists of those hierarchical models
obtained from $m$ by adding one of its dual generators (i.e., minimal interaction terms not present in the model) or deleting one of its generators (i.e.,
maximal interaction terms present in the model). For details see \citep{Edwards, Dellaportas}. For a given set of variables, the algorithm to find $m$ that
maximizes $P(t_{m}|m)$ is analogous to Algorithm 2.1. We give it here for clarity since the notation has changed somewhat.

\begin{algorithm}
\normalfont
We once again make use of a current list $S$ of models that is updated during the search. Define
\[
S(c)=\left\{ m\in S:P(t_{m}|m)\ge c\max_{m'\in M}P\left(t_{m}|m\right)\right\}
\]
where $c\in(0,1)$. A log-linear model $m\in S$ is called explored if all of its neighbours $m'\in\textrm{nbd}(m)$ have been visited.
\end{algorithm}

\begin{enumerate}
\item Initialize a starting list of models $S$. For each model $m\in S$, calculate and record its marginal likelihood $P\left(t_{m}|m\right)$. Mark $m$ as
unexplored.
\item Let $L$ be the set of unexplored models in $S$. Sample a model $m\in L$ according to probabilities proportional with $P\left(t_{m}|m\right)$ normalized
within $L$. Mark $m$ as unexplored.
\item For each $m'\in\textrm{nbd}(m)$, check if $m'$ is currently in $S$. If it is not, evaluate and record its marginal likelihood $P\left(t_{m'}|m'\right).$
Eliminate the models $S\backslash S(c')$ for some pre-chosen value $0<c'<c.$
\item With probability $q$ eliminate from $S$ the models in $S\backslash S(c).$
\item If all the models in $S$ are explored STOP. Otherwise return to step 2.
\end{enumerate}

The prior distribution we use for the log-linear parameters is the generalized hyper Dirichlet of \citep{Massam} which is the conjugate prior for hierarchical
log-linear models. The marginal likelihood $P\left(t_{m}|m\right)$ for an arbitrary $m\in M$ can be approximated using the
Laplace method. This is in contrast to the marginal likelihood of a regression which, as mentioned above, can be computed exactly. The prior distribution has
two hyper-parameters $\alpha$ and $s$ which have particularly easy interpretations. One can think of $s$ as the marginal cell counts of a fictive contingency
table whose cells contain positive real numbers. Then $\alpha$ can be interpreted as the grand total of this table. The \pkg{genMOSS} package only allows a choice 
for $\alpha$ with the default value being $\alpha=1$. There is usually a lack of prior information so the package takes all the fictive cell counts to be equal.

At the end of the two stage MOSS procedure, we will have a list of regressions and best log-linear models describing the associations among the variables in
each regression. After fitting the log-linear models, we can use model averaging to construct a predictor for the response and assess its accuracy using $k$-fold
cross validation. In \pkg{genMOSS}, the log-linear models are fit by finding the mode of the posterior distribution of the log-linear parameters and the predictive weight 
given to each log-linear model is proportional to the marginal likelihood of the corresponding regression (found in the first stage of the MOSS procedure).

\section{The moving window approach}

A simple alternative to stochastically searching through all combinations of SNPs is to group SNPs together according to the sequence that they appear in a
genetic region (if that information is available). This can help identify particular genetic regions of interest. Defining a window size $\omega$ we can first
group SNPs 1 to $\omega$ together and then SNPs 2 to $\omega+1$ together and so on. The marginal likelihood of the regression of each group of SNPs on the
response can be computed as usual using \ref{2.1}. The aim is to identify those regressions such that the marginal likelihood is the highest. The groups of SNPs
(or genetic regions) contained in these regressions are most associated with the response. A weakness of this approach is the subjective selection of the
window size and the inability to capture multi-SNP effects among SNPs in different groups \citep{Sun, Wu}. At the cost of considerable computation, MOSS
circumvents this weakness by allowing any potential SNP to enter a group. The MOSS and moving window approaches are implemented in the \pkg{genMOSS} package in the 
functions \verb!MOSS_GWAS! and \verb!mWindow! respectively. We demonstrate the use of these functions in the following section.

\section{The genMOSS R package}

In this section, \pkg{genMOSS} will be used to analyze a simulated dataset. This dataset, included in the package itself, was simulated using \proglang{Python} code from the
simuPOP \citep{Peng1} cookbook on \url{http://simupop.sourceforge.net}. It is a sample of 1000 cases and 1000 controls from a fictional but realistic population. It contains the genotype information for 6000 diallelic SNPs (i.e., SNPs with three categories) and the disease status for each individual. 
Two SNPs $\textrm{\ensuremath{g_{1}} = 'rs4991689'}$ and $\textrm{\ensuremath{g_{2}} = 'rs6869003'}$ and a random environmental factor $e$ are associated with the disease $\textrm{\ensuremath{Y} = 'aff'}$. The actual model generating the disease has the form:
\begin{equation}
\textrm{logit}\left(P\left(Y=1|g_{1},g_{2},e\right)\right)=-5+0.4g_{1}+0.4g_{2}+0.4g_{1}g_{2}+0.4g_{1}e+0.4g_{2}e.
\label{4.1}
\end{equation}
See \citep{Peng2} for more information about the dataset.

\subsection{Loading the package and data}

The \pkg{genMOSS} \proglang{R} package is available from CRAN and can be installed and loaded by typing:

\begin{verbatim}
R> install.packages ("genMOSS")
R> library("genMOSS")
\end{verbatim}

Next, the simulated dataset, called simuCC (for simulated case-control study) described above, can be loaded with the command:

\begin{verbatim}
R> data("simuCC")
\end{verbatim}

\subsection{Examples}

To run MOSS on the simuCC dataset we use the function \verb!MOSS_GWAS!:

\begin{verbatim}
R> MOSS_GWAS (alpha = 1, c = 0.1, cPrime = 0.0001, q = 0.1, replicates = 5, 
              maxVars = 3, data = simuCC, dimens = c(rep(3,6000),2), 
              confVars = NULL, k = 2)
\end{verbatim}

The parameters c, cPrime, q, and alpha have been described in Section 2. Replicates is the number of instances the first stage of the MOSS procedure will be
run. The top regressions are culled from the results of all the replicates. The parameter maxVars is the maximum number of variables allowed in a regression
(including the response). The variable data is a data frame containing the genotype information for a set of SNPs. It must be organized such that each row refers to a subject 
and each column to a SNP; the last column in data must be a binary response for each subject. Rows with missing values (i.e., NA's) are ignored. Dimens is the number of possible 
values for each column in the dataset. In our example, this is three except for the case-control status which is binary. The parameter confVars (for confounding variables) is a character vector specifying the names 
of SNPs which, other than the response, will be forced to be in every regression. If no confounding variables are desired, confVars can be set to NULL. Finally, the parameter k specifies the fold for the cross validation. 
If k is NULL then only the first stage of MOSS is carried out. In this example, we used the default values for all the parameters (except for k, which is NULL by default, 
and the parameters data and dimens which, of course, are based on the dataset). The output of the above code is:

\begin{verbatim}
$topRegressions
                       formula logMargLik
1 [aff | rs4491689, rs6869003]  -1362.291

$postIncProbs
   variable postIncProb
1 rs4491689           1
2 rs6869003           1

$interactionModels
                         formula logMargLik
1 [rs4491689,aff][rs6869003,aff]   11471.33

$fits
$fits[[1]]

Call:  "[rs4491689,aff][rs6869003,aff]"

Coefficients:
    (Intercept)       rs44916891       rs44916892             aff1
         6.4699          -2.5404          -6.8309          -0.3837
     rs68690031       rs68690032  rs44916891:aff1  rs44916892:aff1
        -0.9192          -3.3279           0.9645           2.1986
aff1:rs68690031  aff1:rs68690032
         0.6603           0.9969

Degrees of Freedom: 17 Total (i.e. Null);  8 Residual
Null Deviance:      4154
Residual Deviance: 12.43        AIC: 112.6

$cvMatrix
     decision
pheno   0   1
    0 644 356
    1 452 548

$cvDiag
   acc  tpr  fpr  auc
1 59.6 54.8 35.6 60.6

\end{verbatim}

The first section of the output gives the top regressions identified by MOSS. In this case, it is the single regression of the response 'aff' on the SNPs
'rs4491689' and 'rs6869003'. By adding the marginal likelihoods of the regressions in which each SNP appears and then normalizing over all the regressions, we
obtain what we call the posterior inclusion probability (postIncProb) for each SNP. These are a measure of each variable's importance and we see, from the
second section of the output, that MOSS attributes great importance to 'rs4491689' and 'rs6869003' (postIncProb = 1), which are in fact the disease
predisposing SNPs. The top log-linear model, shown in the third section of the output, correctly shows an association between the two SNPs and the disease. 
The fitted model, which uses the glm function in the \pkg{stats} \proglang{R} package, is shown in the subsequent section. The last two sections of the output show the results of the cross validation: 'acc' is the accuracy, 'tpr' is the true postive rate, 'fpr' is the false positive rate, and 'auc' is the area under the ROC curve. The 'auc' is computed using the \pkg{ROCR} \proglang{R} package \citep{Sing} available from CRAN. The cross validation results are mediocore here but this is perhaps to be expected since there is an environmental factor in the disease model \ref{4.1}.

The moving window approach described in Section 3 is implemented in the \verb!mWindow! function. Although the SNPs in the simuCC dataset are not necessarily in
sequence, for illustrative purposes, the function call (with a window size of $\omega=2$) is:

\begin{verbatim}
R> s = mWindow (data = simuCC, dimens = c(rep(3,6000),2), alpha = 1, 
                windowSize = 2)
\end{verbatim}

The first three parameters in the \verb!mWindow! function are the same as for \verb!MOSS_GWAS!. The last parameter, windowSize, is the size of the
moving window. After the above code is run, the variable s contains a data frame with the regression in each window and its corresponding log marginal likelihood. The data frame is sorted in descending order by log marginal likelihood. The top 5 regressions can be seen with the command:

\begin{verbatim}
R> head (s, n = 5)

                         formula logMargLik
1   [aff | rs6722027, rs4491689]  -1377.635
2   [aff | rs4491689, rs4450561]  -1384.047
3    [aff | rs325339, rs6869003]  -1385.142
4    [aff | rs6869003, rs325355]  -1385.950
5   [aff | rs6730761, rs3795958]  -1388.191
\end{verbatim}

From the output, we see that the genetic regions around 'rs4491689' and 'rs6869003' (which are the disease predisposing SNPs) show importance. 

For a diallelic SNP, $X$, it may be that the marginal likelihood, $P(Y|X)$, is higher when the SNP is recoded as binary. Using the coding that maximizes this
marginal likelihood may increase the power. Trinary variables can be recoded as binary in three different ways (or can be left as is). The function
\verb!recode_data! in the \pkg{genMOSS} package finds the optimal coding for each diallelic SNP in a given data frame and returns a revised data frame in the same
order as the original. SNPs that are not diallelic are inserted into the new data frame with the coding unchanged. A vector containing the dimension of each SNP in the revised
data frame is also returned. For the simuCC data described above we can run:

\begin{verbatim}
R> s = recode_data (data = simuCC, dimens = c(rep(3,6000),2), alpha = 1)
\end{verbatim}

The three parameters in the \verb!recode_data! function are the same as for \verb!MOSS_GWAS!. The function returns a list with the recoded data frame,
\verb!s$recoded_data!, and the revised dimension vector, \verb!s$recoded_dimens!. For the simuCC dataset, it turns out that the vast majority of the SNPs are
optimally coded as binary. Nevertheless, similar results are obtained when running the \verb!MOSS_GWAS! and \verb!mWindow! functions on the original and
recoded datasets.  

\section{Conclusion}

In this paper we presented the \pkg{genMOSS} \proglang{R} package which can be used for the Bayesian analysis of GWAS data. The package implements the MOSS procedure of \citep{Dobra2} as well as a simple moving window approach to identify combinations of SNPs associated with a response. 
We demonstrated the use of genMOSS on a small simulated dataset which is included with the package. The package can be downloaded from CRAN.

\bibliographystyle{jss}

\nocite{*}.

\end{document}